# Quantum Machine Learning for Software Supply Chain Attacks: How Far Can We Go?

Mohammad Masum
Analytics and Data Science Institute
Kennesaw State University
Kennesaw, USA
mmasum@kennesaw.edu

Mohammad Nazim
Department of Computer Science
Kennesaw State University
Marietta, Georgia, USA
mnazim@students.kennesaw.edu

Md Jobair Hossain Faruk
Department of Software Engineering
Kennesaw State University
Marietta, Georgia, USA
mhossa21@students.kennesaw.edu

Hossain Shahriar, Maria Valero
Department of Information Technology
Kennesaw State University
Marietta, USA
{hshahria,mvalero2}@kennesaw.edu

Md Abdullah Hafiz Khan
Department of Computer Science
Kennesaw State University
Marietta, USA
mkhan74@kennesaw.edu

Gias Uddin, Shabir Barzanjeh, Erhan Saglamyurek
Electrical and Software Engineering, Institute for Quantum Science and Technology, University of Calgary, Canada
{gias.uddin, shabir.barzanjeh, esaglamy}@ucalgary.ca

Akond Rahman
Department of Computer Science
Tennessee Tech University
Tennessee, USA

Sheikh Iqbal Ahamed
Department of Computer Science
Marquette University
Wisconsin, USA
sheikh.ahamed@marquette.

*Abstract*— **Quantum Computing (QC) has gained immense popularity as a potential solution to deal with the ever-increasing size of data and associated challenges leveraging the concept of quantum random access memory (QRAM). QC promises- quadratic or exponential increases in computational time with quantum parallelism and thus offer a huge leap forward in the computation of Machine Learning algorithms. This paper analyzes speed up performance of QC when applied to machine learning algorithms, known as Quantum Machine Learning (QML). We applied QML methods such as Quantum Support Vector Machine (QSVM), and Quantum Neural Network (QNN) to detect Software Supply Chain (SSC) attacks. Due to the access limitations of real quantum computers, the QML methods were implemented on open-source quantum simulators such as IBM Qiskit and TensorFlow Quantum. We evaluated the performance of QML in terms of processing speed and accuracy and finally, compared with its classical counterparts. Interestingly, the experimental results differ to the speed up promises of QC by demonstrating higher computational time and lower accuracy in comparison to the classical approaches for SSC attacks.**

## I. Introduction

A Software Supply Chain (SSC) attack occurs when a cyber threat actor, who locates and attacks technological vulnerabilities, infiltrates a software vendor's network and employs malicious code to compromise the software [1]. Hence, the customer's data or system become compromised as attackers break in and implant malwares into the software before the vendor distributes it to its customers. As a result, a newly developed software may be compromised from the start. On the other hand, customers system may also become compromised during a patch or hotfix as attackers breach these in the vendors' network [1]. These sorts of assaults affect all users and can have far-reaching effects on software consumers at all levels. Hijacking software updates by infiltrating malwares and compromising open-source code are two techniques, frequently employed by threat actors for executing SSC attacks.

First, most software vendors distribute routine updates to patch bugs and security issues via centralized servers to clients as routinary product maintenance. Attackers can hijack the updates and insert malware into the outgoing update, or alter the update and eventually, control over the software's normal functionality. Therefore, this infiltration may cause major disruptions in crucial industries, including international shipping, financial services, and healthcare [5]. As a result, the detection malware is important to prevent unlawful, illegal, unauthorized attacks or access. Traditional anti-malware systems are not capable to combat newly created sophisticated malware [6, 7]. Hence, there is an increasing need for the solutions of automatic malware detection in order to reduce the risks of malicious activities.

Second, reusing crowd-sourced code snippets (e.g., Stack overflow & GitHub) is common practice among software developers to facilitate and expedite the implementation of software applications. However, due to the potential existence of vulnerabilities in such shared code snippets, an SSC attack may occur by compromising the software before the vendor sends it to their customers, which, in turn, affect all compromised software users. As a result, such vulnerabilities could have far-reaching ramifications for government, critical infrastructure, and private sector software users.

Open-source code environments may contain different Common Weakness Enumeration (CWE) vulnerabilities such as Buffer Overflow, Improper Restriction of Operations within the Bounds of a Memory Buffer, Null Pointer Deference, Use of Pointer subtraction to Determine Size, and Improper Input Validation from the abovementioned datasets [3]. Figure 1 displays an example of vulnerable code snippets- a buffer overflow vulnerability example of Linux kernel due to a logic flaw in the packet processor [4]. In-depth semantic reasoning among different components of the code snippets such as variables and functions, is necessary for detecting the code vulnerability, though the fix is simple. Thus, a potential solution is to manually assess and resolve such vulnerable code snippet. However, manually assessing each code is labor-intensive and time-consuming. Therefore, automatic detection of vulnerabilities is crucial for software security.

In recent years, advancements in Machine Learning (ML) and Deep Learning (DL) have facilitated many successful real-world applications ranging from natural language processing to cybersecurity to cancer diagnosis, while achieving better accuracy and performance. However, training ML and DL algorithms encounter challenges, such as high-cost learning and kernel estimation, due to several restrictive factors, including enormous data increase in software supply chain, current computational resources, and high demand to deliver real-time solutions [2]. Meanwhile, Quantum Computing (QC) has gained immense popularity among researchers all over the world as a near-future solution for dealing with the massive amount of data and associated challenges leveraging the concept of quantum random access memory (QRAM) [11]. This paradigm leads to the field of Quantum Machine Learning (QML), promising to overcome the limitations of classical ML and DL. Encoding classical data for QC is an important step in Quantum State preparation and has a significant impact on the overall design and performance of the QML [12]. For instance, amplitude encoding is one of the encoding techniques that requires only $O(\log d)$ qubits in comparison to $O(d)$ bits for classical computing- leading to an exponential compression in the representation of data, which is considered to be the premise for speedup in the quantum version of the methods in Table 1 [12].

Table 1: Time Complexity Analysis for Quantum and Classical Computing

| Methods | Classical Computing | Quantum Computing |
| --- | --- | --- |
| Fast Fourier Transformation (FFT) | $O(d \log d)$ | $O((\log d)^2)$ |
| Eigenvectors and Eigenvalues | $O(d^3)\ O(sd^2)$ | $O((\log d)^2)$ |
| Matrix Inversion | $O(d \log d)$ | $O((\log d)^2)$ |

```
1   static void eap_request(
2       eap_state *esp, u_char *inp, int id, int len) {
3       ...
4       if (vallen < 8 || vallen > len) {
5           ...
6           break;
7       }
8   /* FLAW: 'rhostname' array is vulnerable to overflow.*/
9   -   if (vallen >= len + sizeof (rhostname)){
10  +   if (len - vallen >= (int)sizeof (rhostname)){
11          ppp_dbglog(...);
12          MEMCPY(rhostname, inp + vallen,
                sizeof(rhostname) - 1);
13          rhostname[sizeof(rhostname) - 1] = '\0';
14          ...
15      }
16      ...
17  }
```

Figure 1: Buffer overflow vulnerability in Linux point to point protocol daemon (PPPD)

Therefore, in this study, we explore the promises of QML in comparison to classical ML approaches primarily in the cybersecurity space for malware detection and source code vulnerabilities analysis. We demonstrate a comparative analysis by applying SVM and NN as well as their Quantum version QSVM, and QNN on two real-world datasets: ClaMP dataset for Malware classification and Reveal dataset for source code vulnerability detection.

The rest of the paper is organized as follows: In Section II, we introduce Quantum Machine Learning and existing quantum simulators. Section III includes related work of Quantum Machine Learning. Section IV describes the methodologies: Quantum Neural Network and Quantum Support Vector Machine along with the framework that are implemented in this paper. The experimental setting and results are explained in Section V. Finally, Section VI concludes the paper.

## II. Quantum Machine Learning

Data is stored with Boolean bits at the lowest level in classical computing, where each bit can take only one of two possible values (0 or 1) depending on the existence of electron charge: the existence of electron charge indicates 1, otherwise 0 [13]. On the other hand, the basic unit in quantum computing is quantum bit, referred to Qubit, that can take both the values 0 and 1 simultaneously. Mathematically, qubit state is a vector in two-dimensional (Hilbert) space, described by the linear combination of the two basis states ($|0\rangle$, and $|1\rangle$) in a quantum system: $|\psi\rangle = \alpha|0\rangle + \beta|1\rangle$, where $\alpha, \beta \in \mathbb{C}$ are probability amplitudes that need to satisfy $|\alpha|^2 + |\beta|^2 = 1$ [14]. A qubit state corresponding to such combination of basis states is also called quantum superposition. Furthermore, two qubits can have certain correlations via a quantum phenomenon known as entanglement, which does not have a classical counterpart. When two qubits are entangled, their quantum state cannot be described independently of the state of others. These main principles of quantum machines (superposition and entanglement), give quantum computers enormous power in handling and manipulating many quantum states simultaneously (quantum parallelism), as well as the potential to solve problems that are considered unsolvable in classical computation- leading towards the notion of quantum supremacy [15, 16].

The supremacy of QC promises quadratic or exponential increases in computational time with quantum parallelism for only certain classes of problems. The computation of machine learning algorithms is one of these problems that QC promises to deliver a huge leap. Therefore, in this study, we explored speed up performance of QC when combined with machine learning, known as Quantum Machine Learning (QML). In addition, we investigated comparative analysis of QML and their counterparts classical machine learning in terms of computational time and accuracy. Based on the availability of algorithms both in quantum and classical domains, we selected two existing QML algorithms which are the quantum version of traditional methods: Quantum Support Vector Machine (QSVM), and Quantum Neural Network (QNN).

Executing QML requires access to quantum computers, which unfortunately are rare devices. However, we can leverage publicly available open-source QC frameworks such as IBM Qiskit, TensorFlow Quantum from Google, Amazon's AWS Bracket, Q# and Azure Quantum from Microsoft, and Pennylane from Xanadu that provide simulators to run QML on classical computer. Due to the limitations the state-of-the-art quantum devices and lack of sufficiently large number of qubits, we applied selected QML on the simulator platforms: IBM Qiskit for QSVM and TensorFlow Q for QNN.

IBM Qiskit (Quantum information software kit) is a free and open-source IBM's quantum software development framework, consists of four parts: QASM- operates at the hardware level, Terra- low-level API allows the formation of quantum gates, Aqua- higher-level API that supports machine learning, and Aer- high performance simulator for quantum circuits. Although IBM offers free IBM cloud for computing quantum circuits, waiting time in the queue on the server is extremely long and comes with limited number of qubits (approximately 5 qubits) [10]. On the other hand, Qiskit local simulator comes with much faster processing power as well as a higher number of qubits.

TensorFlow Quantum (TFQ), an extension of open-source python framework Google Cirq, is used for developing QML applications. TensorFlow Quantum integrates with TensorFlow and allows the construction of quantum datasets, quantum models, and classical control parameters as tensors in a single computational graph. In addition, TFQ maintains native integration with the core TensorFlow, principally with Keras model and optimizers. This integration delivers more options towards developing neural network-based architectures, including hybrid quantum-classical neural networks.

## III. Related work

Big data processing requires huge amounts of time, and its classification suffers from this limitation as well, rendering quantum computing based classification a suitable option to manage such data [18, 19, 20]. One of the explored quantum-inspired classifications is the Quantum Least Square Support Vector Machine (Quantum LS-SVM) [18]. Quantum LS-SVM has average values and standard deviations of classification rates of 91.45 % in low-rank datasets and 89.82% in low-rank approximate datasets while the classical computer's Library for Support Vector Machine (LIBSVM) have 86.46% and 84.90% classification rates respectively. Furthermore, implementation on a quantum computer utilizing a quantum big data algorithm (i.e., non-sparse matrix exponentiation for matrix inversion of the training data inner-product matrix) and quantum evaluation can be done directly in a higher-dimensional space using a quantum kernel machine [20]. Another approach is Quantum Multiclass SVM, which is based on quantum matrix inversion algorithm and one-against-all strategy. This approach maps datasets to quantum states, and uses of QRAM for accessing data in quantum parallel, and finally, performs memory access incoherent quantum supposition, which results in quadratic speed gain in comparison to existing approaches in classical computers [19].

Binary classification on remote sensing (RS) of multispectral images can be achieved on D_WAVE 2000Q Quantum Annealer machine using Quantum SVM [17]. This method formulates the classification as a Quadratic Unconstrained Binary Optimization (QUBO) and implements the RBF kernel method on two ID datasets: Im16, and Im40. The method achieved AUROC score of 0.886 and AUPRC score of 0.930 for Im16 dataset, respectively. AURCOC of 0.882 and AURPC of 0.870 were achieved for the other dataset Im40, respectively [17]. Similar RS testing for image classification on 50 samples from SemCity Toulouse dataset on an upgraded quantum machine- D-WAVE Advantage- produced an overall accuracy of 0.874 with 0.734 F1 score which were comparable to classical SVM models and outshone the IBM quantum machines that lagged with 0.609 and 0.569 scores respectively [21]. QSVM with RBF kernel and SVM (the classical counterpart) were applied to the Wisconsin breast cancer dataset [24]. The QSVM was implemented on Qiskit aqua with a real backend quantum-chip (ibmqx4) and obtained an accuracy of 80%, whereas the classical SVM performed better with an accuracy of 85%. However, the study found that using QSVM on a simulator surpassed the traditional approach by reaching near-perfect accuracy. The same study conducted a Quantum multiclass variational SVM on the UCI ML Wine dataset to achieve 93.33% accuracy on the iqmqx4 and 100% accuracy on StateVector simulator while the local CPU environment can reach 90% accuracy with classical SVM.[8]. Quantum neural networks (QNN) was applied to various datasets, including Fisher's Iris dataset, modified Iris dataset, Sonar dataset, and Wisconsin's Breast Cancer dataset, using the single-shot training scheme, which allows input samples can be trained in a single qubit quantum system [22]. The QNN producing accuracy of 83.26%, 96.96%, 41.25% and 90.19%, respectively, outperforming a classical NN with zero hidden layer [22]. However, when two more hidden layers were added to architecture, the classical NN outperformed the QNN.

In the application of Field-programmable gate arrays (FPGAs), a data structure, referred to as n-BQ-NN which contains the learning framework of n-bit QNNs can attain an almost exact accuracy of full-precision models while being 68.7% energy efficient and 2.9 times higher performance than SVPE (shift-vector processing element) by replacing multiply operations with SHIFT operations on ResNet, DenseNet, and AlexNet network structures [25]. Additionally, a variation of Grover's quantum search algorithm (called BBHT optimization), finds the optimal weights of a neural network and train a QNN more efficiently for data classification [23]. This model is constructed by stimulation of a Perceptron with a step activation function when the first qubit of the inner product result of input and weights of neuron is measured by the usage of quantum Fourier transformation [23].

Dynamic traffic routing can be determined by extracting live data from devices on GPS landmarks which are preprocessed in Spark SQL and further processed by a combination of Decision tree and Random Forest before being fed to QNN to accurately show the best route from a specific source to destination [26]. Testing accuracy of QNN with single hidden layer of 97.3%, 97.5% and 85.5% for corresponding training pairs of 75, 30,12 respectively on Iris Dataset which was comparable to both classical neural networks of CVNN and RVNN with single hidden layer [27]. However, the computational speed for QNN ran 100 epochs were as compared with CVNN ran for 1000 epochs and RVNN for 5000. Furthermore, quantum feature maps based on Quantum Random Access Coding (QRAC) has been used on Variational Quantum Classifiers (VQC) that resulted in better performance and efficiency by utilizing small number of qubits on Breast Cancer (BC) dataset and Titanic Survival (TS) dataset with a test accuracy and f1 score of 0.682 and 0.483 for BC and 0.772 and 0.707 for TS dataset [28]. Earth Observation (EO) dataset called EuroSat had CNN and QNN4EO (QNN for EO), which formed of three convolutional 2D layers used for image classification showed QNN4EO and reached an accuracy of 94.73%, outperforming the 93.63% accuracy of CNN [29].

## IV. METHODOLOGY

We applied classical ML classifiers such as Support Vector Machine and Neural Network and their quantum versions- Quantum Support Vector Machine (QSVM) and Quantum Neural Network (QNN), respectively. We implemented the methods on two SSC attack datasets: ClaMP and ReVeal. Figure 2 displays the framework describing the process of implementation. After collecting the raw data, data pre-processing techniques were used to prepare the data to input to the methods. In the preprocessing step for ClaMP data: categorical data were converted into numerical features and later all the features were normalized to maintain a similar scale. In the preprocessing step for ReVeal data: each of the code snippet were embedded into an identical sized vector using some pre-trained model. Since the accessibility to large number of quantum bits is limited, we reduced the dimension of both datasets. On one hand, the reduced data is directly input to the classical version of the classifiers. On the other hand, the reduced features were encoded into quantum states before feeding to the quantum classifiers: QSVM and QNN.

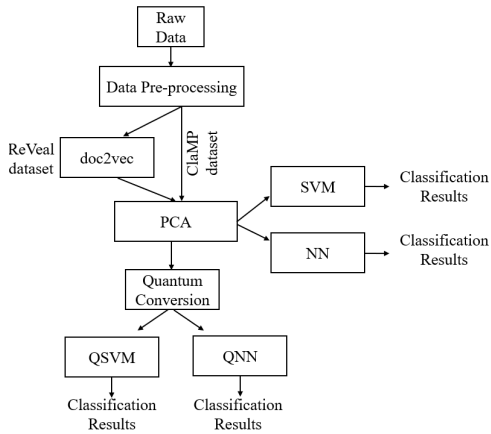

Figure 2: Architecture of the Framework

Quantum Neural Network (QNN) emerges from the theory of neurocomputing that intersect novel concepts including machine learning algorithm, quantum computing, and artificial neural networks [30]. Considering the size, depth, and precision complexity, QNN framework can be applied for vast levels of information processing capacity of neural computing that can provide enormous potential in solving various combinatorial optimization problems.

The input data is encoded into the relevant qubit state of an appropriate number of qubits, and the Quantum Neural Network (QNN) processes it [31]. The qubit state is then modified for a specified number of layers using parameterized rotation gates and entangling gates where the predicted value of a Hamiltonian operator, (for instance- Pauli gates), is used to determine the altered qubit state. These results are decoded and converted into useful output data. An optimizer, such as Adam optimizer, then updates the parameters while a Variational Quantum Circuits (VQC)-based neural network plays a variety of functions in many forms in quantum neural networks (QNN). The complexity-theoretic measurements of size, depth, and accuracy characterize distinct features of computations where the number of steps, requiring to solve an issue is measured in depth. The size of the equipment typically corresponds to the magnitude of the problem; precision also describes the apparatus required to solve the problem.

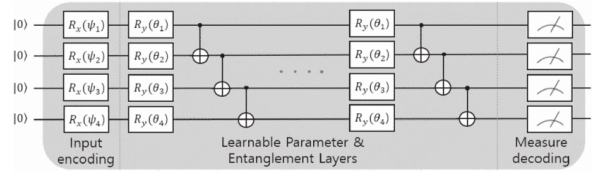

Figure 3: depicts the QNN with the input parameter and linear entanglement structure [31]

A quantum neural network consists of input, output, and L hidden layers. Quantum circuit of quantum perceptron is structured into L hidden layers of qubits that acts on an initial state of the input qubits and produces an, in general, a mixed state for the output qubits. QNNs' ability to do universal quantum computation, even for two-input one-output qubit perceptron, is a direct result of the quantum-circuit construction that considers quantum perceptron on 4-level qubits. The most generic version of the quantum perceptron may implement any quantum channel on the input qubits. The precision of $p(n)$ is denoted by $\{s(n), d(n)\}$, whereas size and depth are denoted by $s(n)$ and $d(n)$, respectively, which are created from the gates D and U of precision $p(n)$. The number of qubits in the circuit is measured in size, while the longest sequence of gates from input to output is measured in depth. To eliminate the problem of localization, the reversible U gate is usually followed by the irreversible D gate. The accuracy of the circuits is usually $O\{s(n)\}$.

Quantum Support Vector Machine (QSVM) is a high-performance version of an important machine learning technique that provides data privacy advantages and utilizes as a component in a larger quantum neural network [32, 33].

QSVM can be adopted for training data to classify complex problems and a quantum computing processor has the potential of conducting experiments in larger datasets than those of current computing system. Such advancement is due to more qubits and higher connectivity (up to 15 connections per qubit, instead of up to 6 connections per qubit) which pave to classify experiments with a QSVM implementation on the quantum circuit model. In both the training and classification stages, a quantum support vector machine can be developed with a various run times, including O (log NM) [33]. Binary classification problems can be addressed using QSVM where various methods can be applied including variational method and the quantum kernel-based method [34].

The primary advantage of quantum variational approach is that it can process multiple classification for the response variable while requiring to run two sequential quantum algorithms that lead to more computationally intensive than the quantum kernel-based method. After the support vectors have been created with a classical computer, classification may begin to predict the labels for the test data set using the conventional computing. By adopting the QSVM approach, different methods are used to train data and estimate the result with the quantum computer.

## V. Experiments and results

### A. Dataset specification

We applied ML algorithms: SVM and NN as well as their Quantum version QSVM, and QNN on two real-world datasets: ClaMP dataset for Malware classification and Reveal dataset for source code vulnerability detection.

There are two versions of ClaMP: 1. ClaMP_Raw- contains only raw features and 2. ClaMP_Integrated- contains both raw and extracted features. We used the ClaMP_Integrated version. The raw malware samples were collected from VirusShare, while the benign samples were collected from Windows files. From both malware and benign samples, features were collected from Portable Executable (PE) headers, since the PE header contains all the required information that OS needs to run executables. Additionally, the PE header contains useful information regarding malware functionality and the interactive nature between malware and OS. Thus, several raw features (55 features) were extracted using the rule-based method from PE headers of the samples including DOS header (19 features), File Header (7 features), and Optional Header (29 features. Meaningful features are derived using raw features such as entropy, compilation time, section name, etc. In addition, a set of raw features were expanded from the File header to extract more information about the PE file. Finally, a set of raw, derived, and expanded features were selected to form the ClaMP_Integrated dataset, containing in total 68 features, where the number of raw, expanded, derived features are 28, 26, and 14, respectively [8].

ReVeal is a real-world source code dataset where vulnerabilities are tracked from Linux Debian Kernel and Chromium open-source projects [9]. Large evolutionary history, program domains containing diverse security issues, and publicly available vulnerability reports made the dataset a more robust and comprehensive compared to other existing datasets in source code vulnerability such as STATE IV, SARD, and Draper datasets. Readily fixed issues with publicly available patches were collected using Bugzilla for Chromium and Debian security tracker for Linux Debian Kernel. Vulnerability issues associated with each of the patches were identified by filtering out commits that do not have security related keywords. The dataset contains a vulnerable version (annotated as vulnerable) of C/C++ source and header file as well as the fixed version (annotated as clean) corresponding to the vulnerable version. In addition, other functions, not involved with the patch were remained unchanged and annotated as a clean source code. Figure 4 displays an example of such data collection process [9], where two versions of func.c (version k-1 and version k) are included. The red function ham_0 in the previous version (version k-1) was fixed to ham_1 function. The dataset would contain both versions with annotating ham_0 vulnerable and ham_1 as non-vulnerable code snippet. Other two functions: spam() & egg() would remain unchanged and labeled as non-vulnerable.

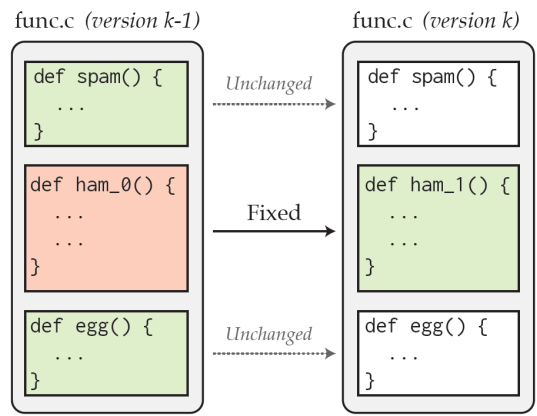

Figure 4: ReVeal Data collection process: Green indicates non-vulnerable code, while red indicates vulnerability [9].

The ReVeal dataset contains a total of 22,734 samples, with 2240 non-vulnerable and 20,494 vulnerable samples, respectively. We randomly selected 2240 samples without replacement from the non-vulnerable samples to balance the dataset.

### B. Data Preprocessing

We applied SVM, NN and their quantum counterparts QSVM, and QNN on ClaMP and ReVeal datasets. We vary the size of the data to examine the comparative performance of the methods when a lesser size of data is used. For the ClaMP dataset, we first considered the entire dataset, which included

5210 samples. Thereafter, we randomly selected 75 percent and 50 percent of the dataset without replacing any samples and constructed two smaller datasets with 3907 and 2605 samples, respectively, while preserving the class proportion. Similarly, we created two smaller datasets from the ReVeal dataset, with 3360 and 2240 samples, respectively, encompassing 75% and 50% of the original dataset. We divided the six datasets into 70 percent training data and 30 percent test data, with techniques being trained on the training and evaluated on the test datasets, respectively.

Categorical values cannot be entered directly into the model. The ClaMP data comprises one categorical variable, 'packer type,' which was converted into a numerical variable while avoiding the dummy variable trap by removing one random category from all of them. As a result, (40-1) = 39 dummy variables were added to the dataset, resulting in a total of 108 columns including one target variable. Because the features in the dataset are on different scales, we used a normalizing approach (standardization) to transform all the features to the same scale with a mean of zero and a standard deviation of one. In addition, to avoid data leakage issues, we fit the standardization technique to the training data and then transform both the training and test data.

## C. Experimental Setting

Doc2Vec model was applied to the samples of the ReVeal dataset for converting the text into a numerical vector of size 100. We set the window size (maximum distance between the current and predicted word within a sentence) to 10, alpha (initial learning rate) to 0.01, minimum alpha (linear decay rate) to 0.0001, minimum count (ignore all words with total frequency lower than a threshold) to 2 and epoch 50. We used a vector size of 100 to capture more context from the data. However, the present quantum simulator cannot accept such a dimension as an input. As a result, we used another dimension reduction strategy on this.

A dimension reduction technique, Principal Component Analysis (PCA), was applied to the vector of size 100 of the ReVeal dataset and to the 108 features of the CLaMP dataset for reducing the dimension. Due to the limitation of qubit numbers in the existing simulator, we selected first 16 principal component that contains 98%, 99%, and 75% of the variation of the three datasets, respectively. The classical SVM and NN were directly applied to all the reduced datasets.

Next step is to encode the classical data as quantum circuits, i.e., converting each of the features' value into qubit for further processing it in the quantum computer or simulator. Figure 5 displays the circuit created for a random sample. These circuits (Cirq) were then converted into TFQ. Next, we developed model circuit layer for the QNN (Figure 6). We built a two-layer model, matching the data-circuit size and finally wrapped the model-circuit in a TFQ-Keras model, where the converted quantum data were fed, and Parametrized Quantum Layer (PQC) was used to train the model circuit on the quantum data. In training, hinge loss was used as an optimization function. Thus, we converted the labels to [-1, 1]. Finally, we trained the QNN for 20 epochs. We applied classical neural networks-based architecture containing single and multiple hidden layers to compare the results with QNN, where 51, and 177 parameters were included in the single- and two-hidden layers classical NN. We applied the single hidden layer NN to offer a fair comparison to the QNN. In addition, we developed two hybrid QNN models (Hybrid-QNN_V1 & Hybrid-QNN_V2), each containing 45 and 63 parameters, respectively. The hybrid models contain one PQC and one classical Keras layer, where the Keras layer contains 4 and 10 nodes in the hidden layer for Hybrid-QNN_V1 & Hybrid-QNN_V2, respectively.

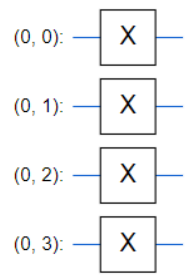

Figure 5: Conversion to quantum data point

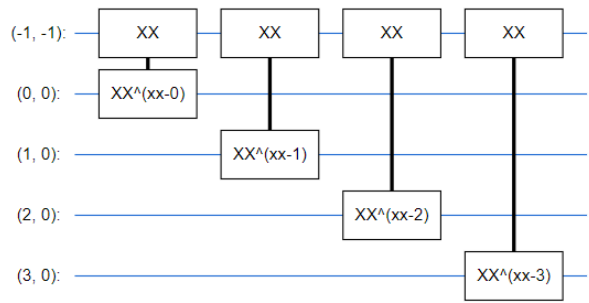

Figure 6: An example of circuit layer of Quantum Neural Network

For ClaMP dataset, we followed a similar pre-processing step: PCA was applied, and 16 reduced features were fed to the QNN classifier as well as its' classical version NN containing two hidden layers with 177 parameters. In addition, a classical NN-fair version was applied, including one hidden layer with 73 parameters for providing a fair comparison between the QNN and NN.

*D. Results*

Table 1 displays Comparative results analysis of Classical and Quantum Machine Learning Classifiers on CLaMP DataSet. Both the Quantum classifiers perform poorly in comperison to their counterparts classical approaches in terms of accuracy and total execution time. The QNN produces only 52.1 accuracy, while taking a large amount of execution time: 2698 seconds. On the other hand, Classical NN and Classical NN-Fair version produces much higher accuracy 92.7% and 90.5%, respectively, while taking extremely lower execution time: 22 and 19 seconds, respectively. Support Vector Machine, as well, shows similar patters in terms of accuracy and total time. The QSVM provides 73.5% accuracy, whereas the classical SVM provides 93.5% accuracy. The QSVM consume significantly higher execution time (10000 seconds) than the Classical SVM, showing the efficiency of classifcal computing.

Table 2: Comparative results analysis of Classical and Quantum Machine Learning Classifiers on Entire CLaMP DataSet

| Model | Parameters | Accuracy (%) | Time (s) |
|---|---|---|---|
| QNN | 32 | 52.1 | 2698 |
| Hybrid-QNN_V1 | 45 | 52.27 | 2581 |
| Hybrid-QNN_V2 | 63 | 52.27 | 2507 |
| Classical NN | 177 | 92.7 | 22 |
| Classical NN-Fair | 73 | 90.5 | 19 |
| QSVM | | 73.5 | 10000 |
| Classical SVM | | 93.5 | 1 |

The application of QML on the ReVeal dataset demonstrates the ineffectiveness by producing significantly lower performance in terms of execution time, though the accuracy provided by both systems is approximately similar. All versions of QNN, including the Hybrid methodologies, provide 52.71% accuracy, while the execution time is considerably higher than the classical counterparts. The quantum versions took approximately 60-fold longer times to execute the program compared to the classical methods. Similarly, the classical SVM outperformed the QSVM both in terms of accuracy and speed, though difference in accuracy was not significant enough. The SVM achieved 60.34 percent accuracy with a very short execution time, whereas the quantum version achieved 58.26 percent accuracy with a significantly longer execution time (16682 seconds). The lower performance of the simulated quantum computing may be attributed to the limited number of qubits for producing better accuracy and accessibility of open-source quantum simulators.

Table 3: Comparative results analysis of Classical and Quantum Machine Learning Classifiers on Entire ReVeal Dataset

| Model | Parameters | Accuracy (%) | Time (s) |
|---|---|---|---|
| QNN | 32 | 52.71 | 3006 |
| Hybrid-QNN_V1 | 45 | 52.71 | 2999 |
| Hybrid-QNN_V2 | 63 | 52.71 | 2563 |
| Classical NN | 177 | 55.7 | 41 |
| Classical NN-Fair | 51 | 52.74 | 20 |
| QSVM | - | 58.26 | 16682 |
| Classical SVM | - | 60.34 | 2 |

*E. Discussion*

QML has limitations because its applicability is entire dependent on quantum hardware, and quantum hardwire (simulator) necessitate a considerable amount of computational capacity to study a large number of events and qubits. In addition, the number of quantum operations often limited by the increasing errors from decoherence that can be performed on a noisy quantum computer [35]. This was evident in our analysis, as we had long queuing time and execution time with a larger number of observations. The time required to initialize qubits and measure them in the current simulator may result in a lengthy execution time for QML algorithms. Furthermore, due to the simulator's constraint of existing qubits, we used a limited qubit, which may result in poor performance for the QML methods. This raises the important question of how many qubits are required to exhibit quantum advantages in the analysis of software supply chain attacks.

Although there are limitations of current quantum computing and accessibility in quantum devices, this study shows that QML can leverage high dimensionality of quantum state space to deal with real world big cybersecurity data.

## VI. CONCLUSION

Quantum Computing (QC) has gained immense popularity among researchers and promised to deliver a huge leap forward in the computation of Machine Learning algorithms. This paper analyzes speed up performance Quantum Machine Learning such as Quantum Support Vector Machine (QSVM), and Quantum Neural Network (QNN) to detect software supply chain attacks. The QML methods were applied on open-source quantum simulators such as IBM Qiskit and TensorFlow Quantum. We evaluated the performance of QML in terms of processing speed and accuracy. The experimental results differ to the speed up promises of QC by producing significantly lower accuracy and taking higher execution time in comparison to their classical counterparts. Though the QC has the potential to revolutionize computation, current versions with limited number of qubits are not advanced enough to produce rewarding performance, specifically, in software supply chain attacks. However, QML algorithms that use an advanced quantum computer or simulator with a large number of qubits may surpass their classical machine learning equivalents in terms of classification performance and computational time.